\documentclass[aps,prl,twocolumn,nopacs,superscriptaddress]{revtex4-1} 

\usepackage{amsmath}
\usepackage{bbm}
\usepackage{epsfig}
\usepackage{color}

\usepackage{times}

\newcommand{\cc}{{\mathbbm{C}}}

\begin{document}

\title{Cooling by heating}
\author{A.\ Mari} 

\affiliation{Institute for Physics and Astronomy, University of Potsdam, 14476 Potsdam, Germany}
\affiliation{Dahlem Center for Complex Quantum Systems, Freie Universit{\"a}t Berlin, 14195 Berlin, Germany}

\author{J.\ Eisert}
\affiliation{Institute for Physics and Astronomy, University of Potsdam, 14476 Potsdam, Germany}
\affiliation{Dahlem Center for Complex Quantum Systems, Freie Universit{\"a}t Berlin, 14195 Berlin, Germany}

\begin{abstract}
We introduce the idea of actually cooling quantum systems by means of incoherent thermal
light, hence giving rise to a counter-intuitive mechanism of ``cooling by heating''. In this effect, 
the mere incoherent occupation of a quantum mechanical mode serves as a trigger to enhance the
coupling between other modes.
This notion of effectively rendering states 
more coherent by driving with incoherent thermal quantum noise is applied here to the opto-mechanical setting, 
where this effect occurs most naturally. We discuss two ways of describing this situation,
one of them making use of stochastic sampling of Gaussian quantum states with respect to 
stationary classical stochastic processes. The potential of experimentally
demonstrating this counter-intuitive effect in opto-mechanical systems with present technology is 
sketched.
\end{abstract}

\maketitle

Cooling in quantum physics is usually achieved in just the same way as it occurs in classical physics or in common
everyday situations: One brings a given system into contact with a colder bath. Coherent driving of quantum systems
can effectively achieve the same aim, most prominently in instances of laser cooling of ions or in its opto-mechanical variant, 
cooling mechanical degrees of freedom using the radiation pressure of light. The coherence
then serves a purpose of, in a way, rendering the state of the system ``more quantum''. 
In any case, in these situations, the interacting body should first and foremost be cold or coherent. 


In this work, we introduce a paradigm in which thermal hot states of light can be used to 
significantly cool down a quantum system. To be specific, we will focus on an opto-mechanical
\cite{Aspe,Schliesser,Groeblacher,Vitali,VedralVitali,Review} implementation of this idea: This
type of system seems to be an ideal candidate to demonstrate this effect with present
technology; it should however be
clear that several other natural instances can well be conceived. 
Intuitively speaking, it is demonstrated that due to the driving with thermal noise, the interaction of other modes can be effectively 
enhanced, giving rise to a ``transistor-like'' effect \cite{Remark}.
We flesh out this effect at hand of two approaches following different approximation schemes. 
The first approach is essentially a weak coupling master equation, while the second approach makes 
use of stochastic samplings with respect to colored classical stochastic processes 
\cite{Kampen}, which constitutes an interesting and practical tool to study such quantum
optical systems of several modes in its own right.

The observation made here adds to the insight that appears to be appreciated only fairly recently,
in that quantum noise does not necessarily only give rise to heating, decoherence, and dissipation,
providing in particular a challenge in applications in quantum metrology and 
in quantum information science. When suitably used, quantum noise can 
also assist in processes thought to be necessarily of coherent nature, in 
noise-driven quantum phase transitions \cite{Zoller}, quantum criticality \cite{Prosen}, in 
entanglement distillation \cite{Distillation} or in quantum computation \cite{Verstraete}. It turns out that 
thermal noise, when appropriately used, can also assist in cooling. Alas, this counterintuitive effect
is not in contradiction to the laws of thermodynamics, as is plausible 
when viewing this set-up as a thermal machine or heat engine \cite{Popescu}  operating in the quantum regime. 

\begin{figure}[tbh]
\centerline{\includegraphics[width=0.34\textwidth]{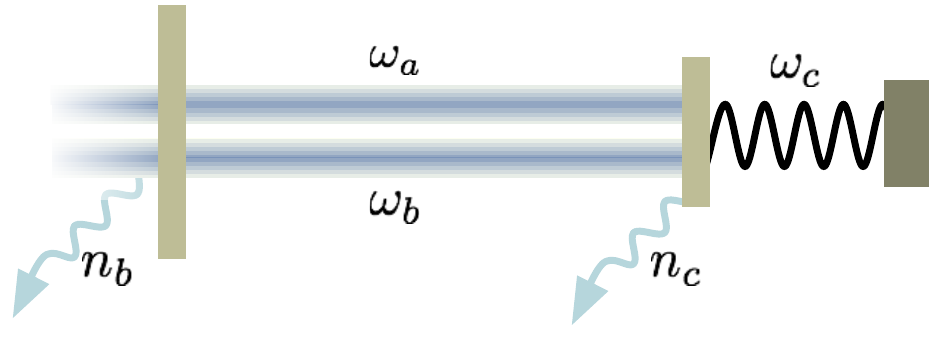}} 
\caption{The opto-mechanical setup primarily being considered in this work, involving two optical modes and a mechanical one.} \label{system}
\end{figure}

{\it The system under consideration.}
We consider a system of two optical modes at frequencies $\omega_a$ and $\omega_b$, respectively,
that are coupled to a mechanical degree of freedom at frequency $\omega_c$.
The Hamiltonian of the entire system is assumed to be well-approximated by
$ H=H_0+H_{1}$,
where the free part is given by
 $H_0=\hbar \omega_a a^\dag a + \hbar \omega_b b^\dag b+ \hbar\omega_c c^\dag c$, 
and the interaction can be cast into the form
\begin{equation}
  H_1=\hbar g (a+b)^\dag(a+b) (c+c^\dag).  \label{h1}
\end{equation}
It is convenient to move to a rotating interaction picture with respect to $\hbar \omega_b(a^\dag a+b^\dag b)$. 
The radiation pressure interaction is invariant under this transformation, while $H_0$ simplifies to
\begin{equation*}
 H_0'=\hbar \Delta a^\dag a + \hbar\omega_c c^\dag c,
\end{equation*}
where $\Delta=\omega_a-\omega_b$. For most of what follows, the frequencies are chosen such that 
$\Delta=\omega_c$, 
as we will see is the optimal resonance for cooling the mechanical resonator.
This can be realized by tuning the mechanical degree of freedom or the cavity mode splitting. 
In fact, this is exactly the setting proposed in Ref.\ \cite{Yin}
as a feasible three-mode optoacoustic interaction, in an idea that can be traced back
to studies of parametric oscillatory instability in Fabry-Perot interferometers \cite{Brag}.
Similarly, with systems of high-finesse optical cavities coupled to thin semi-transparent membranes 
\cite{Membrane}, of double-microdisk whispering-gallery resonators \cite{Painter} or
of opto-mechanical crystals \cite{Crystal} such a situation can be achieved. 
Surely numerous other architectures are well conceivable. 

In addition to this coherent dynamics, the system is assumed to undergo natural damping and decoherence -- unavoidable
in the opto-mechanical context.
The quantum master equation governing the dynamics of the entire system embodying the two optical
modes and the mechanical degree of freedom
is given by
\begin{equation}
  \dot \rho=\mathcal L \rho= -\frac{i}{\hbar}[H,\rho]+ (\mathcal L_a+\mathcal L_b+\mathcal L_c)\rho,  \label{master}
\end{equation}
with the generators being defined by $\mathcal L_a =\kappa D_{a}$ and
\begin{eqnarray}
\mathcal L_b &=&  (1+n_b)\kappa D_{b}+n_b \kappa D_{b^\dagger}, \\
\mathcal L_c&=&  (1+n_c)\gamma D_{c}+n_c \gamma D_{c^\dagger}, \label{lc}
\end{eqnarray}
making use of the notation
for a generator  in Lindblad form
\begin{equation}
D_x (\rho) =2 x\rho x^\dag-\{x^\dag x, \rho\} . \label{dissipator}
\end{equation}
Here, we allow the optical bath of mode $b$ to be in a Gibbs or thermal state having an arbitrary temperature.

This type of damping reflects the plausible mechanism of loss. For the mechanical motion, we are primarily interested in
the regime where
$\omega_m\gg \gamma$, such that the damping mechanism of quantum Brownian motion based on some spectral density
is virtually indistinguishable from the quantum optical Markovian damping as for an optical mode \cite{Check}. For that reason,
for coherence of presentation, the same type of dissipative dynamics has been chosen for the optical and mechanical modes.

We will now discuss this given situation in two different pictures. The first one is a weak coupling approach leading to 
approximate analytical expressions. The second one involves sampling over colored classical stochastic processes.
These methods are further discussed in the range of their validity in the EPAPS, where they are also compared with
exact diagonalization methods for small photon numbers \cite{EPAPS}.

{\it Description 1: Weak coupling approximation as an analytical approach.} In this approach, a picture is developed
grasping the physical situation well for small couplings $g$. In addition to the actual physical baths 
of the three modes
$a$, $b$, and $c$ giving rise to dissipative dynamics, we also consider mode $b$ as
a further external ``bath'' and derive an effective master equation for modes $a$ and $c$ only.
This is a good approximation if the back action on mode $b$ is negligible and up to 
second order in the coupling constant $g$.
Having this picture in mind, the Liouvillian in Eq.\ (\ref{master}) can be decomposed as 
$\mathcal L=\mathcal L_{\rm sys}+\mathcal L_{\rm int}+\mathcal L_{\rm bath}$, where
$\mathcal L_{\rm bath}=\mathcal L_b$ and
\begin{eqnarray}
\mathcal L_{\rm sys}=-\frac{i}{\hbar}[H_0', \cdot ]+\mathcal L_a+\mathcal L_c ,\,\,\,
\mathcal L_{\rm int}=-\frac{i}{\hbar}[H_1, \cdot ] \label{Lint} .
\end{eqnarray}
Using projection operators techniques \cite{Gardiner}, 
one can derive a master equation for the reduced system 
$\rho_{a,c}=\textrm{tr}_{b}[ \rho]$
\begin{equation*}
	\dot \rho_{a,c}(t)= \mathcal L_{\rm sys}\rho_{a,c}(t)+
	\textrm{tr}_b \mathcal L_{\rm int} \int_0^\infty ds \,  e^{\mathcal L_r s} \mathcal L_{\rm int} \rho_{a,c}(t-s) \otimes \rho_b. 
\end{equation*}
Here $\mathcal L_r=\mathcal L_{\rm sys}+\mathcal L_{\rm bath}$.
Making use of the explicit expression (\ref{Lint}) for $\mathcal L_{\rm int}$, we have
\begin{equation}
	\dot \rho_{a,c}(t)= \mathcal L_{\rm sys}
	\rho_{a,c}(t)-\frac{1}{\hbar^2} \textrm{tr}_b  [H_1, \int_0^\infty ds  
	e^{\mathcal L_r s} [H_1, \rho_{a,c}(t-s)\otimes \rho_b] ] \label{masterH}
\end{equation}
In what follows, we will make a sequential approximation
 of the interaction Hamiltonian $H_1$ and the damping mechanism. 
In order to be as transparent as possible, we mark each of the steps with a roman letter.

Eq.\ (\ref{masterH}) -- up to 
second order expansion in the coupling $g$, which constitutes the first approximation step (a) -- can also 
be written as
\begin{equation}
	\dot \rho_{a,c}(t)= \mathcal L_{\rm sys}\rho_{a,c}(t)-\frac{1}{\hbar^2}\textrm{tr}_b  [H_1, 
	\int_0^\infty ds  [ e^{\mathcal L_r^\dag s} (H_1), 
	\rho_{a,c}(t)\otimes \rho_b] ], \label{masterint}
\end{equation}
where $\mathcal L_r^\dag$ acts only on the Hamiltonian $H_1$, 
corresponding to a ``dissipative interaction picture" with respect to $\mathcal L_r$.

We start from Eq.\ (\ref{h1}) and (b) neglect the term proportional to $a^\dag a$ because we assume mode $a$ to be weakly perturbed from 
its ground state. In contrast, we allow the physical optical bath of mode $b$ to have an arbitrary 
temperature and therefore we cannot neglect the term proportional to $b^\dag b$. 
We rewrite the approximated $H_1$ as
\begin{equation}
 H_1'=\hbar g (a^\dag b+b^\dag a+\delta) (c+c^\dag), \label{h1prime}
\end{equation}
where the operator $\delta=b^\dag b-n_b $ represents the intensity fluctuations of mode $b$.
In order to have vanishing first moments with respect to mode $b$, 
the mean force proportional to $\langle b^\dag b \rangle$
has been subtracted, 
which is responsible of merely
shifting the resonator equilibrium position. 
Since $\omega_a-\omega_b=\omega_c$, 
the (c) rotating wave approximation (RWA) of Eq.\ (\ref{h1prime}) is 
\begin{equation}
 H_1''=\hbar g (a^\dag b c+a b^\dag c^\dag)+\hbar g\delta (c+c^\dag). \label{h1second}
\end{equation}
As will be explained later in more details, the first term of the Hamiltonian is responsible for the   
\textit{cooling} of the mechanical resonator, while the second term corresponds to 
an additional \textit{heating} noise.

In order to compute the partial trace in Eq.\ (\ref{masterint}), we need the two-time correlation
functions of the thermal light in mode $b$,
\begin{eqnarray}
\langle b e^{\mathcal L_r^\dag s}b^\dag \rangle=e^{-\kappa s}n_b, \,\,\,\, \label{corr}
\langle \delta e^{\mathcal L_r^\dag s}\delta \rangle=e^{-2\kappa s}(n_b^2+n_b). 
\end{eqnarray}
The exponential functions in Eqs.\ (\ref{corr}) determine the time scale of the integral kernel in 
Eq.\ (\ref{masterint}), which will be of the order of $\kappa^{-1}$.
Within this time scale (d) 
we can neglect the effect of the mechanical reservoir ($\gamma\ll\kappa$), and the action of the map $e^{\mathcal L_r^\dag s}$ on the system operators will be 
\begin{eqnarray*}
e^{\mathcal L_r^\dag s} a&=&e^{-(\kappa+i \Delta) s}a=e^{-(\kappa+i \omega_c) s}a, \\
e^{\mathcal L_r^\dag s} c&=&e^{-(\gamma+i\omega_c) s} c\simeq  e^{-i \omega_c s} c. 
\end{eqnarray*}
We can finally perform the integral in Eq.\ (\ref{masterint}), and since all 
the odd moments of $\rho_b$ vanish, the \textit{cooling} and \textit{heating} terms in 
Eq.\ (\ref{h1second}) generate two independent
contributions to the master equation, respectively 
\begin{eqnarray}
 \mathcal L_{\rm cool} &=&\frac{g^2}{2 \kappa} \left((1+n_b) D_{a c^\dag}+n_b D_{a^\dag c}\right), \label{lcool} \\
\mathcal L_{\rm heat} &=& \frac{2 \kappa g^2 (n_b^2+n_b)}{4 \kappa^2+\omega_c^2}\left(D_{c^\dagger}+D_{c}\right),\label{lheat} 
\end{eqnarray}
where in calculating $\mathcal L_{\rm heat}$ we (e) kept only the counter-rotating terms.
The effect of $\mathcal L_{\rm heat}$ is simply a renormalization of the mean occupation number of the mechanical bath
\begin{equation*}
n_c \mapsto \tilde n_c= n_c+ \frac{2 \kappa g^2 (n_b^2+n_b)}{\gamma (4 \kappa^2+\omega_c^2)},
\end{equation*}
always increasing, as expected,  the effective temperature of the environment.
Denoting with $\tilde{\mathcal L}_{\rm sys}$ the corresponding renormalized Liouvillian, the
master equation can be written as
\begin{equation}
\dot \rho_{a,c}=( \tilde{\mathcal L}_{\rm sys}+\mathcal L_{\rm cool} )\rho_{a,c}.  \label{master2}
\end{equation}
With respect to Eq.\ (\ref{master}), Eq.\ (\ref{master2}) can be numerically solved with much less computational 
resources but we have to remind ourselves
that this approach is valid only within the RWA and for weak coupling:  $\gamma, g \ll \omega_c $.
Another advantage of Eq.\ (\ref{master2}) is that the corresponding adjoint equations for the 
number operators $n_a=a^\dag a$ and $n_c= b^\dag b$ are closed with respect to these operators, that is
\begin{eqnarray*}
\dot{n}_a  &=&
- 2 \kappa  n_a-\frac{g^2}{\kappa}\left( (n_b+1) n_a
-n_b  n_c
 -  n_a  n_c\right) , \\
\dot{n}_c 
&=&
- 2 \kappa n_c-\frac{g^2}{\kappa}\left( n_b n_c
-(n_b+1) n_a
 - n_a n_c\right)
 +2\gamma \tilde n_c \nonumber .
\end{eqnarray*}
Assuming (f) that the factorization property $\langle n_a n_c \rangle\simeq\langle n_a\rangle\langle n_c \rangle$ holds -- which is 
essentially a mean-field approach which is expected to be good in case of small correlations, or, again as assumed, 
for small values of $g$ -- we can 
find analytical expressions for the steady state expectation values:
\begin{eqnarray*}
\langle n_c\rangle &=& \frac{\tilde n_c-\eta}{2}
+\left({ \frac{(\tilde n_c+\eta)^2}{4}-\frac{\kappa n_b \tilde n_c }{\gamma}}\right)^{1/2}, \\
\langle n_a\rangle &=& \frac{(\tilde n_c-\langle n_c\rangle) \gamma}{\kappa},
\end{eqnarray*}
where
$\eta=1 + n_b(1+ \kappa/\gamma) + 2 \kappa^2/g^2$.

{\it Description 2: Sampling with respect to colored stationary classical stochastic processes.}
In this approach, 
we start from the exact dynamics Eq.\ (\ref{master}) but treat mode $b$ as a classical thermal field and neglect any feed-back from the resonator.
We substitute the bosonic operator with a complex amplitude $b(t) \mapsto \beta_t$, giving rise to a semi-classical picture.
The parameter $\beta_t$ can be described as a classical stochastic process defined 
by the stochastic differential equation (SDE)
\begin{equation}
 d\beta_t= -\kappa \beta_t dt + \sqrt{\kappa n_b} (dW^{(x)}+idW^{(y)})\label{dbeta}, 
\end{equation}
with independent Wiener increments \cite{Kampen}
obeying the It\=o rules $dW^{(a)}dW^{(b)}=\delta_{a,b} dt$, $dW^{(a,b)} dt=0$.
The dynamics of the remaining modes $a$ and $c$ instead, can be efficiently treated quantum mechanically; this is true,
since for every single realization of the process (\ref{dbeta}), the evolution defines a Gaussian completely positive 
map and therefore the corresponding Gaussian state $\rho_{a,c}^{(\beta_t)}(t)=\mathcal E_t^{(\beta_t)} (\rho_{a,c})$ 
can be described entirely in terms of first and second moments.
The actual quantum state of the system will in general not be exactly 
Gaussian, it can nonetheless be simulated by sampling over many Gaussian states associated with different realizations of $\beta_t$: Only the
respective weight in the convex combinations are such that the resulting state can be non-Gaussian.
The resulting state $\rho_{a,c}(t)=\mathbbm E\rho_{a,c}^{(\beta_t)}(t)$ will be our semi-classical description of the system.

\begin{figure}[tbh]
\centerline{\includegraphics[width=0.35 \textwidth]{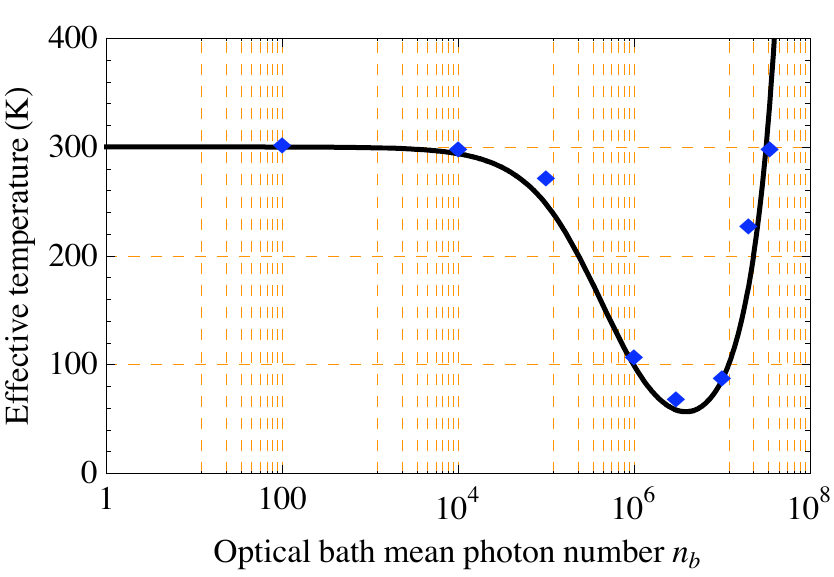}} 
\caption{Room temperature cooling with parameters reminding of those typical in realistic 
experiments \cite{Groeblacher}: $\omega_c= 2\pi$MHz, $\kappa= 0.2\omega_c$, $g= 0.3 \times 10^{-5}\omega_c$, and $\gamma= 10^{-3}\omega_c$. 
The black line shows the predictions of the steady state using Description 1, the dots are a result from stochastic sampling using
Description 2 (with $100$ realizations), which qualitatively coincide well. One clearly finds that an increased population of mode $b$ leads to a significant
cooling of the mechanical mode -- up to a point when eventually the ``bad noise'' becomes dominant. 
} \label{ch}
\end{figure}

It is convenient to introduce a vector of quadratures operators $u=[x_c,y_c,x_a,y_a]$, where $x_j=(j+j^\dag)/\sqrt{2}$,
$y_j= i(j^\dag-j)/\sqrt{2}$ and $j=a,c$.
From Eq.\ (\ref{master}), we get a SDE for the first moments
\begin{equation}
\frac{d \langle u\rangle_t}{dt} =A_t \langle u \rangle_t+f_t  , \label{du}
\end{equation}
where 
\begin{eqnarray}
  A_t=\left[\begin{array}{cccc}
    	-\gamma  & \omega_c	& 0 		& 	0 \\
     	-\omega_c & -\gamma & g\beta^{(x)}_t & g\beta^{(y)}_t  \\
        -g\beta^{(y)}_t	  & 	0	& -\kappa 	&  \Delta \\
    	 g\beta^{(x)}_t   &     0	& -\Delta & -\kappa
  	\end{array}\right]  \hspace{-0.4 em},
	f_t=\left[\begin{array}{cccc}
    	0 	   \\
     	 g|\beta_t|^2\\
          0\\
    	 0
  	\end{array}\right] \label{af}
\end{eqnarray}
$\beta_t^{(x)}=(\beta_t+\beta^*_t)$, $\beta_t^{(y)}=i(\beta^*_t - \beta_t)$.
The second moments can be arranged in the matrix $V_t={\rm Re} \langle u u^\dag\rangle_t$,
satisfying the SDE
\begin{equation}
\frac{dV_t}{dt}=A_t V_t+V_tA_t^T+D+F_t, \label{dv}
\end{equation}
where $D=\textrm{diag}[\gamma(2n_c+1), \gamma(2n_c+1), \kappa, \kappa]$,
and $F_t=f_t \langle u \rangle_t^T+ \langle u \rangle_t f_t^T$.
The statistical average over many realization of $V_t$ will be an estimator for 
the second moments of the quantum state $V(t)=\mathbbm{E} (V_t)$. In particular, the first two diagonal elements give the effective 
phonon number of the mechanical oscillator, since  $\langle  n_c \rangle(t)=(V_{1,1}(t)+V_{2,2}(t)-1)/2$.
The three stochastic differential equations (\ref{dbeta},\ref{du},\ref{dv}) can be numerically integrated in sequential order. In our simulations,
see Fig.\ \ref{ch}, we 
used the Euler method, for each time step $dt$ sampling 
the associated Wiener increments in Eq.\ (\ref{dbeta}) with normal distributions of variance $\sigma^2=dt$.

{\it Intuitive explanation of the effect of cooling by heating.} This effect can be intuitively explained at hand of Eq.\ 
(\ref{h1second}) in Description 1: Two competing processes play here an important role:
 The first term appears like a beam splitter interaction between the modes $a$ and $c$ with a ``reflectivity" 
given by the thermal fluctuations of the amplitude of mode $b$. This is responsible for the cooling of the mirror. 
That is to say, the occupation of mode $b$ takes the role resembling the ``basis of a transistor'': A high occupation
renders the interaction between $a$ and $c$ stronger, hence triggering the cooling effect. For this effect to be 
relevant, the coherence or purity of the state of $b$ does not play a dominant role, and hence even
thermal noise can give rise to cooling. This is referred to as ``good noise''.
The second term corresponds to the fluctuations of the radiation pressure of mode $b$ and it is a source of  
``bad" noise which heats the mechanical mode.

Similarly, this effect can be studied at hand of the stochastic picture of Description 2, when 
observing Eq.\ (\ref{dv}). In addition to the intrinsic quantum noise described by $D$, stochastic fluctuations of $\beta_t$ generate an 
additional heating noise given by the matrix $F_t$. However, the same process $\beta_t$ is also contained in the matrix $A_t$ and  
corresponds to a cooling noise, up to the approximations identical to the above ``good noise''.
The reason is quite evident from Eq.\ (\ref{af}), where we observe that the coupling between the hot 
mechanical oscillator and the cold optical mode is mediated by the thermal fluctuations of $\beta_t$. 
This opto-mechanical coupling, which would be zero 
without noise, leads to a sympathetic cooling of the mechanical mode.

{\it Example.} We will now discuss the effect of cooling by heating at hand of an example using 
realistic parameters in 
an opto-mechanical setting. Fig.\ \ref{ch} shows the effective temperature of the mechanical mode
as a function of the number of photons in mode $b$: Here, effective temperature
is defined as the temperature $T$ of a Gibbs state 
\begin{equation*}
	\rho_c(T) = \frac{e^{-\hbar \omega_c c^\dag c/(kT)}}{\textrm{tr} (e^{-\hbar \omega_c c^\dag c/(kT)})}
\end{equation*}
such that $\langle  n_c \rangle = {\textrm{tr}} (\rho_c(T) c^\dagger c)$. One quite impressively
encounters the effect of cooling by heating, for increasing photon number and hence effective temperature of this optical mode. 
For very large values of the photon number, the ``bad noise'' eventually becomes dominant,
resulting again in a heating up of the mechanical mode. Note that needless to say, the effective temperature of the optical
mode $b$ is usually larger than the mechanical one by many orders of magnitude (approximately $10^{10}$K for reasonable parameters).
  
{\it Summary.} In this work, we have established the notion of cooling by heating, which means
that cooling processes can be assisted by means of incoherent hot thermal light. We focused
on an opto-mechanical implementation of this paradigm. We also introduced
new theoretical tools to grasp the situation of driving by quantum noise, including sampling techniques
over stochastic processes. 
To experimentally demonstrate this counterintuitive effect should be exciting in its own right. 
Putting things upside down, one could also conceive settings similar to the one discussed here
as demonstrators of small heat engines \cite{Popescu} operating at the quantum mechanical level,
where $b$ takes the role of an ``engine'' and mode $a$ of a ``condenser''. To fully explore these
implications for feasibly realizing quantum thermal machines constitutes an exciting perspective.
It would also be interesting to fully flesh out the potential for the effect to assist in generating non-classical states
\cite{Modulation}. Finally, quite intriguingly, this work may open up ways to think of 
optically cooling mechanical systems without using lasers at all, but rather with basic, cheap LEDs 
emitting incoherent light. 


{\it Acknowledgements.} We would like to thank the EU (Minos, Compas, Qessence), the 
EURYI, and QuOReP for support.

\section{Supplementary material}

In this supplementary material, we compare the methods of Description 1 and 2 of the main text with an exact
simulation of the master equation (\ref{master}) in a truncated, finite-dimensional
Hilbert space of the three involved modes, $  \mathcal H=\cc^{N^{a}}\otimes \cc^{N_b}\otimes \cc^{N_c}$.
The unique stationary state of Eq.\ (\ref{master})
can easily be found numerically; a dimension $d=N_a N_b N_c$ of the total Hilbert space of, say, $d \lesssim 400$, 
is well feasible. This is obviously an essentially exact method for small occupation numbers in each of the 
three modes, and the error made can easily be estimated. This analysis, see Fig.\ 3, together with analogous ones
in similar regimes, shows that the methods used here are also suitable in the deep quantum regime.

\begin{figure}[tbh]
\centerline{\includegraphics[width=0.42 \textwidth]{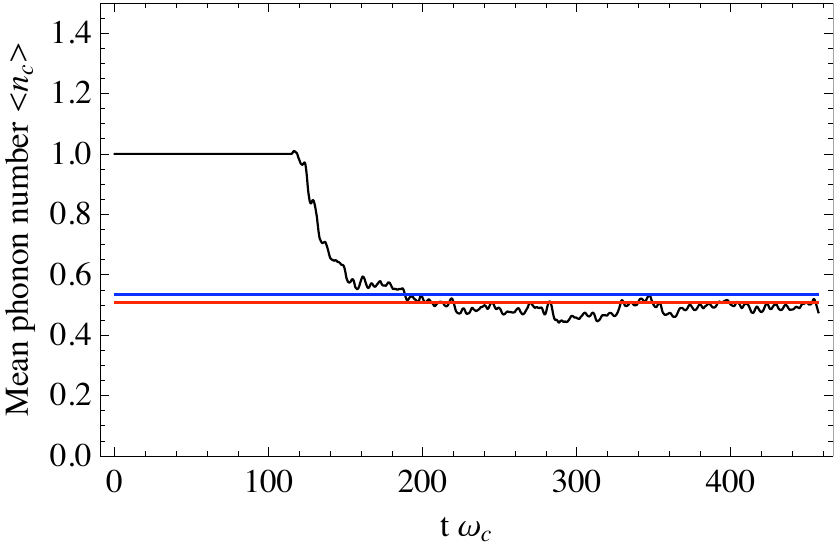}} 
\caption{Stochastic simulation -- introduced in Description 2 --  of the mean number of phonons as a function of time (black line), compared with the exact steady state (blue line) and with the analytical approximation given in Description 1 (red line). Parameters: $n_b=n_c=1$, $\kappa=0.1 \omega_c$, $\gamma=0.01 \omega_c$, $g=0.006 \omega_c$.} \label{sim1}
\end{figure}



\begin{thebibliography}{99}
\bibitem{Remark}
	This cooling is not due to some energy minimum being favored, with a transition facilitated
	by small amounts of quantum noise, reminding of quantum versions of stochastic resonance \cite{SR}.
\bibitem{Check}
	A picture of quantum Brownian motion has also been compared to this
	quantum optical Markovian limit, with no notable differences in any predictions within the 
	parameters that are of interest in this work.
\end{thebibliography}
\end{document}